\documentclass[a4paper]{jpconf}

\usepackage{pb-diagram}
\usepackage{amsmath,amsfonts,amssymb,amsmath,epsfig}
\usepackage[english]{babel}
\usepackage{graphicx}
\usepackage{bm}
\usepackage{tikz-cd}
\usepackage[hidelinks,linktoc=all]{hyperref}
\usepackage{cite}
\usepackage{physics}
\usepackage{tabularx}
\usepackage{comment}

\def\cD{{\cal D}}

\newcommand{\be}{\begin{equation}}
\newcommand{\ee}{\end{equation}}

\def\br{\begin{eqnarray}}
\def\er{\end{eqnarray}}

\def\({\left(}
\def\){\right)}
\def\[{\left[}
\def\]{\right]}

\def\lie{{\cal G}}
\def\a{\alpha}

\def\l{\lambda}

\def\pa{\partial}

\def\tp0{\Thet\mathcal{A}_{+}^{(0)}}
\def\tm0{\Thet\mathcal{A}_{-}^{(0)}}

\def\l{\lambda}

\def\bi{\begin{itemize}}
\def\ei{\end{itemize}}

\newcommand{\kdv}{\text{KdV}}
\newcommand{\mkdv}{\text{mKdV}}

\newcommand{\dI}{\partial_x^{-1}}

\DeclareMathAlphabet{\mathpzc}{OT1}{pzc}{m}{it}

\renewcommand{\u}{\mathcal{J}}
\renewcommand{\v}{\mathpzc{V}}

\begin{document}

\title{Complex KdV  rogue waves from \\gauge-Miura transformation}
\author{Ysla F. Adans,
Guilherme  Fran\c ca,
Jos\' e F. Gomes,
Gabriel V. Lobo, and
Abraham H. Zimerman
}
\address{Institute of Theoretical Physics --- IFT/UNESP,
Rua Dr. Bento Teobaldo Ferraz 271, 01140-070, S\~ ao Paulo, SP, Brazil}
\ead{
ysla.franca@unesp.br,
guifranca@gmail.com,
francisco.gomes@unesp.br,
gabriel.lobo@unesp.br,
a.zimerman@unesp.br}

\begin{abstract}The gauge-Miura correspondence establishes a map between the entire
KdV and mKdV hierarchies, including positive and also negative flows,
from which new relations besides the standard Miura transformation arise.
We use this correspondence to classify solutions of the 
KdV hierarchy in terms of elementary tau functions of the mKdV hierarchy
under both zero and  nonzero vacua.
We  illustrate how interesting nonlinear phenomena can be described
analytically from this construction, such as  
``rogue waves'' of a complex KdV system that corresponds to a limit of 
a vector nonlinear Schr\" odinger equation.
\end{abstract}

\section{Introduction}

Recently, the seminal Miura transformation was extended to a gauge
transformation between all differential equations of the
KdV and mKdV hierarchies \cite{Adans_2023}.
In this paper we review some of these results and explore these connections 
to systematically classify 
KdV solutions in terms of  mKdV tau functions.
This not only allow us to put several important integrable models
on a common framework but also to construct interesting solutions thereof.
For instance,
we obtain various solutions of the KdV hierarchy 
such as solitons, dark solitons, peakons, kinks, breathers,
negatons, and positons from their  mKdV counterparts.  Moreover, 
we consider complex solutions of the mKdV and KdV
hierarchies under a zero and nonzero background.
Some of these solutions correspond to ``rogue waves,'' i.e., 
extreme waves of abnormal amplitude, having great  interest
in diverse fields such as ocean waves, optics, plasma physics, 
and Bose-Einstein
condensates
\cite{Solli_2007,Kibler_2010,Baronio_2012,Tsai_2016,Dudley_2019}.

\section{KdV and mKdV hierarchies}
\label{sec:kdv_mkdv}

\subsection{Positive flows}
The KdV,
\be\label{kdv0}
4 \pa_{t_3} \u - \pa_x^3 \u + 6 \u  \pa_x \u = 0 ,
\ee
and mKdV,
\be \label{mkdv0}
4 \pa_{t_3} \v - \pa_x^3 \v + 6 \v^2 \pa_x \v = 0 ,
\ee
equations are two important nonlinear integrable models. They 
played a fundamental role in the development of the inverse
scattering transform whereby the \emph{Miura transformation},
\be \label{miura0}
\u = \v^2 + \pa_x \v ,
\ee
is key.
Such a relation generates solutions of the KdV equation from
solutions of the mKdV equation. Conversely, it is
a Riccati differential equation for $\v$ given $\u$.
Moreover, the Miura transformation also allows one to obtain a Schr\" odinger
spectral problem and construct an infinite number of conserved charges of the
KdV equation \cite{Gardner_1967}.

The KdV and mKdV equations turn out to be a single  model inside a more general
structure, namely an \emph{integrable hierarchy} of differential equations.
An integrable hierarchy contains an infinite set of differential equations
whose flows are in \emph{involution}. Integrable hierarchies
can be systematically
constructed from the \emph{zero curvature} condition
\cite{Zakharov_1972,Drinfeld_1985},
\be
\label {zero_curvature}
[ \pa_x + A_x, \pa_{t_N} + A_{t_N} ] = 0 ,
\ee
with gauge potential $A_x$ and $A_{t_N}$ taking values  on 
an affine (Kac-Moody) Lie algebra with a suitable grading structure. Importantly, 
the operator $A_x$ uniquely defines the hierarchy, and for each admissible $N$,
eq.~\eqref{zero_curvature} generates one particular flow (differential equation).

For instance, the affine Lie algebra $\widehat{s\ell}(2)$ has generators
$E_{\pm \alpha}^{(n)} \equiv \lambda^{n} E_{\pm\alpha}$ and
$h^{(n)} \equiv \lambda^{n} h$, where 
\be
\big[ E_{\alpha}, E_{-\alpha} \big] = h, \qquad
\big[ h, E_{\pm \alpha}\big] = \pm 2 E_{\pm \alpha}
\ee
are the generators of the (finite-dimensional)  Lie algebra $s\ell(2)$. 
Here $\lambda \in \mathbb{C}$ is the so-called
spectral parameter. The 
\emph{principal grading operator} $\widehat{Q} \equiv  2 \tfrac{d}{d\lambda} + \tfrac{1}{2} h^{(0)}$  decomposes $\lie = \widehat{s\ell}(2)$ into 
$\lie_{2n} = \big\{ h^{(n)}\big\} $ (grade $2n$) 
and 
$\lie_{2n+1} = \big\{ E_{\alpha}^{(n)} + E_{-\alpha}^{(n+1)} \big\}$ (grade $2n+1$).
Moreover $\mathcal{K}_E = \lie_{2n+1}$ is the kernel of 
the semi-simple element $E \equiv E_{\alpha}^{(0)} + E_{-\alpha}^{(1)}$,
which has odd degree.
Thus, the \emph{positive part of the mKdV hierarchy} can be constructed with
\begin{subequations}
\begin{align}
\label{Ax_mkdv}
A_x^{\mkdv} &\equiv E + \v h^{(0)}, \\ 
A_{t_N}^{\mkdv}  &\equiv D^{(N)}_N + D^{(N-1)}_N + \cdots+ D^{(0)}_N 
\qquad \big(D^{(n)}_N \in \lie_n \big) . 
\label{At_mkdv}
\end{align}
\end{subequations}
Since the kernel $\mathcal{K}_E$ has only odd elements it follows that
the allowed positive flows are odd,
$N= 1,3,5,\dotsc$.
For a given $N$, the consistency of 
the zero curvature condition \eqref{zero_curvature}
determines $A_{t_N}$  in terms of the field $\v$ and its derivatives, 
ultimately leading to a  differential equation associated 
to time $t_{N}$.
For instance, the  mKdV equation \eqref{mkdv0} is obtained with $N=3$.
Another example with $N=5$ is the modified Sawada-Kotera equation,
\be \label{mSK}
16 \pa_{t_5} \v  - \pa_x^5 \v  + 40 \v \pa_x \v \pa_x^2 \v + 10 \v^2 \pa_x^3 \v
+ 10 (\pa_x\v)^3 - 30 \v^4 \pa_x\v  = 0 ,
\ee
and one can obtain higher order differential equations for $N=7,9,\dotsc$.
On the other hand, the \emph{positive part of the 
KdV hierarchy} is constructed with
\begin{subequations}
\begin{align} 
\label{Ax_kdv}
A_x^{\kdv} &\equiv E +  \u E_{-\a}^{(0)}, \\
A_{t_N}^{\kdv} &\equiv  
{ \cal D}^{(N)}_N+ {\cal D}^{(N-1)}_N + \cdots + {\cal D}^{(0)}_N + {\cal D}^{(-1)}_N \qquad \big({\cal D}^{(n)}_N \in \lie_n\big). 
\label{At_kdv}
\end{align}
\end{subequations}
Again, the grading operator $\widehat{Q}$ only allows $N$  \emph{odd}.
The KdV equation \eqref{kdv0} is obtained with $N=3$,
the Swada-Kotera equation
\be \label{SK}
16 \pa_{t_5}\u - \pa_x^5\u + 20 \pa_x \u \pa_x^2 \u + 10 \u \pa_x^3 \u - 30 \u^2 \pa_x \u = 0 
\ee
with $N=5$, and so on for $N=7,9,\dotsc$.

It is possible to extend the 
the Miura transformation \eqref{miura0} to the entire KdV and mKdV hierarchies;
we call this correspondence \emph{gauge-Miura transformation} \cite{Adans_2023}.
Since the zero curvature eq.~\eqref{zero_curvature} 
is invariant under gauge transformations, we can require
\be
\label{gauge}
\begin{split}
A_x^{\kdv} = S  A_x^{\mkdv} S^{-1} + S \pa_x S^{-1}
\end{split}
\ee
and attempt to solve for $S$. As it turns out, there are two solutions:
\be
S_{1} = \begin{pmatrix}
  1 & 0\\ 
  \v & 1
\end{pmatrix} , \qquad 
S_{2} = \begin{pmatrix}
  0 & \l^{-1}\\ 
  1 & -\l^{-1}\v
\end{pmatrix} ,
%=  
%\begin{pmatrix}
%  1 & 0 \\ 
%  -\v & 1 
%\end{pmatrix} \lambda^{-1} E 
\label{miura_gauge}
\ee
such that 
\be \label{miura}
\u =  \v^2  \mp \pa_x \v ,
\ee
respectively.
Thus, the Miura transformation arises as a ``symmetry'' condition
between the entire KdV and mKdV hierarchies ---
it can also be shown that
$A_{t_N}^{\kdv} = S A_{t_N}^{\mkdv} + S \pa_x S^{-1}$ is automatically
satisfied. Hence, the gauge-Miura transformation  
connects all positive flows of the KdV and mKdV hierarchies in
a 1-to-1 fashion; 
replacing an mKdV solution $\v(x, t_{N})$ into eq.~\eqref{miura} generates
two different KdV solutions $\u(x, t_{N})$:
\be\label{corresp_pos}
\begin{tikzcd}[row sep=0.1em, column sep=3em]
t_{N}^{\mkdv} \arrow[r, "S"] & t_{N}^{\kdv} .
\end{tikzcd}
\ee

Finally, we stress the role of the \emph{vacuum} which is a trivial
yet important solution of these equations.
Note that $\v_0 = 0$ solves both eqs.
\eqref{mkdv0} and \eqref{mSK}, and so does a \emph{constant} $\v_0 \ne 0$.
The same is true for eqs. \eqref{kdv0} and \eqref{SK}, i.e., $\u_0 = 0$ and
$\u_0 = \mbox{const.} \ne 0$ are solutions;
it can be shown from the zero curvature equation and the grade
structure of these hierarchies that this remains true for all positive
flows of the mKdV and KdV hierarchies. 
Each vacuum generates an orbit of  solutions via the action of dressing operators.  The differential equations 
within the positive
parts of the KdV/mKdV hierarchies simultaneously 
admit solutions over zero or nonzero vacua.

\subsection{Negative flows}

The mKdV and KdV hierarchies can be extended to incorporate
negative  flows, which often (but not always) correspond to integro-differential equations. The operators \eqref{Ax_mkdv} and \eqref{Ax_kdv}
are fixed, however the temporal gauge potentials are now
required to have the form
\begin{align}
A_{t_{-N}}^{\mkdv}  &\equiv 
D^{(-N)}_{-N}+ D^{(-N+1)}_{-N}+\cdots D^{(-1)}_{-N} \qquad \big(D^{(-n)}_{-N} \in \lie_{-n}\big) , 
\label{At_mkdv_neg}
\\
A_{t_{-N}}^{\kdv} &\equiv \cD^{(-N-2)}_{-N} + \cD^{(-N-1)}_{-N} + \cdots + \cD^{(-1)}_{-N} \qquad \big( \cD^{(-n)}_{-N} \in \lie_{-n}\big) 
\label{At_kdv_neg}
.
\end{align}
With these gauge potentials we can solve 
eq.~\eqref{zero_curvature}.  
For the mKdV hierarchy, one obtains with  
$-N = -1$ the sinh-Gordon model
($\v = \pa_x \phi$),
\be 
\pa_{t_{-1}} \pa_x \phi =  2 \sinh ( 2 \phi ) ,
\label{sg_eq}
\ee
while for  $-N = -2$ one obtains 
\be  \label{mkdv_minus_two_eq}
\pa_{t_{-2}}   \pa_x \phi = - 2 e^{-2 \phi} \dI e^{2 \phi} - 2 e^{2 \phi} \dI e^{-2 \phi} ,
\ee
where $\dI f \equiv \int f  dx$ such  
that $\pa_x \pa_x^{-1} = \pa_x^{-1} \pa_x = 1$. 
Importantly,  for the negative part of the mKdV hierarchy 
there is a differential equation associated to each integer
$-N = -1,-2,\dotsc$, i.e., \emph{odd or even}, which is in
contrast with the postive part.

For the negative part of the KdV hierarchy
only negative \emph{odd} flows are admissible.
For $-N=-1$ one obtains
($\u \equiv \pa_x \eta$)
\be 
\pa_{t_{-1}} \pa_x^3  \eta - 4 \pa_x\eta \pa_{t_{-1}} \pa_x  \eta
- 2  \pa_x^2 \eta \, \pa_{t_{-1}} \eta = 0 .
\label{kdv_neg1}
\ee
This equation, which was first proposed in
\cite{Verosky_1991}, is thus the KdV counterpart of the sinh-Gordon model.  
It is also possible to obtain other  models for
$-N = -3,-5,\dotsc$. 

The gauge-Miura transformation
\eqref{gauge} also holds true for the negative flows. However,
there is a degeneracy in the sense that two consecutive odd and even mKdV flows
are mapped into the same odd KdV flow:
\be\label{corresp_neg}
\begin{tikzcd}[row sep=-0.8em, column sep=3em]
t_{-N}^{\mkdv} \arrow[dr, "S"] & \\
&  t_{-N}^{\kdv} \\ 
t_{-N-1}^{\mkdv} \arrow[ur, swap, "S"] 
\end{tikzcd} .
\ee
This degeneracy  gives rise to 
new relations \emph{in addition} to the typical Miura transformation \eqref{miura}. For instance, when mapping
the sinh-Gordon \eqref{sg_eq} into eq. \eqref{kdv_neg1} the following
identity arises:
\be \label{mt1}
\pa_{t_{-1}} \eta(x, t_{-1})  = 2  e^{\mp 2\phi(x, t_{-1})}.
\ee
Similarly, in mapping eq.~\eqref{mkdv_minus_two_eq} into eq.~\eqref{kdv_neg1}
such a relation becomes
\be \label{mt2}
\pa_{t_{-1}} \eta(x,t_{-1}) = \pm 4 e^{\mp 2\phi(x,t_{-2})} \dI e^{\pm 2\phi(x, t_{-2})} . 
\ee
The signs follow from $S_1$ or $S_2$ in eq.~\eqref{miura_gauge}, respectively.

There is a peculiarity regarding the \emph{vacuum} structure 
for the negative part of the mKdV hierarchy. 
A zero vacuum $\v_0 = 0$ --- or $\phi_0 = 0$ --- is   a solution
the sinh-Gordon  \eqref{sg_eq}, but not of eq.~\eqref{mkdv_minus_two_eq}.
Conversely, a constant vacuum  
$\v_0 \ne 0$ --- or $\phi_0 = \v_0 x$ --- 
is a solution of eq.~\eqref{mkdv_minus_two_eq}, 
but not of sinh-Gordon \eqref{sg_eq}.
It can be shown that this pattern  is general, i.e., all \emph{negative
odd} flows of  the mKdV hierarchy only admit  \emph{zero vacuum solutions},
 while all the \emph{negative even} flows only admit constant 
 \emph{nonzero vacuum} solutions \cite{Franca_2009}.
However, the situation is different for the KdV hierarchy:
eq. \eqref{kdv_neg1} simultaneously admits
$\u_0 = 0$ --- $\eta_0 = 0$ --- or $\u_0 = \mbox{const.} \ne  0$ 
--- $\eta_0 = \u_0 x + c \, t_{-N} $, $\u_0 = \v_0^2$.
The same behavior generalizes to all \emph{negative  odd} KdV flows.
We can understand this from the gauge-Miura relationship \eqref{corresp_neg}. 
A zero vacuum KdV solution for time $t_{-N}^{\kdv}$ (odd) 
is inherited from a zero vacuum mKdV solution
of time $t_{-N}^{\mkdv}$ (odd),  while
a nonzero vacuum KdV solution for time $t_{-N}^{\kdv}$ (odd)
is inherited from the nonzero vacuum mKdV solution for time 
$t_{-N-1}^{\mkdv}$ (even).

\section{KdV solutions in terms of mKdV tau functions}
\label{sec:tau}

From the above connections, the mKdV hierarchy plays a more fundamental
role than the KdV hierarchy since solutions of the latter can be systematically
constructed from the former. 
\emph{Solutions of the mKdV hierarchy} --- in the orbit of a zero, $\v_0 = 0$, or nonzero, $\v_0 = \mbox{const.} \ne 0$, vacuum ---
can be constructed in terms of
tau functions \cite{Franca_2009}:
\be\label{v_sol}
\v = \pa_x \phi, \qquad \phi = \v_0 x + \log \big( \tau^- / \tau^+ \big). 
\ee
The general form of the tau functions for an  ``$n$-soliton'' is 
($j,k,\ell,\dotsc  = 1,\dotsc,n$)
\cite{Franca_2009}
\be\label{taus}
\begin{split}
\tau^{\pm} &=  1 + \sum_{j} C_j^{\pm} \rho_j
+ \sum_{j < k} C_j^\pm C_k^\pm A_{jk}  \rho_j \rho_k
+ \sum_{j < k < \ell } C_j^\pm C_k^\pm C_\ell^\pm A_{jk} A_{j\ell} A_{k\ell} \rho_j \rho_k \rho_\ell
+ \dotsm \\
& \qquad + 
C_1^\pm \dotsm C_n^\pm \Big( \prod_{j < k} A_{jk} \Big) \rho_1 \dotsm \rho_n ,\end{split}
\ee
where
\be\label{coeffs}
C_{j}^{\pm} = \dfrac{\v_0 \pm \kappa_j}{2 \kappa_j}, \qquad
A_{jk} = \left(\dfrac{\kappa_j - \kappa_k}{\kappa_j + \kappa_k}\right)^2 , \qquad
\rho_j = e^{f_j(x, t_{N})} .
\ee
The function $f_j(x,t_N)$ encodes 
the \emph{dispersion relation}, which depends on the wave number $\kappa_j$,
and is determined up to an arbitrary constant phase.
For zero vacuum all \emph{odd flows} (positive or negative, 
$N=\pm 1, \pm 3,\dotsc$)  have
$f_j(x, t_{\pm N}) = 2 \kappa_j x + 2 \kappa_j^{\pm N} t_{N}$.
For constant vacuum all \emph{negative even flows} 
($N = 2,4,\dotsc$) have
$f_{j}(x,t_{-N}) = 2 \kappa_j x + \tfrac{2 \kappa_j}{\v_0(\kappa_j^2 - \v_0^2)^{N/2}} t_{-N}$. There is no well-defined pattern for the positive odd flows,  but one can find the dependency case-by-case, e.g.,
for $N=3$ we have
$f_j(x,t_3) = 2 \kappa_j x + \big(2 \kappa_j^3 - 3 \v_0^2 \kappa_j\big)t_{3}$.
Importantly, these solutions have the same form for all models within the mKdV  hierarchy:
only the dispersion relation changes in terms of the factor with $t_N$ that
controls the wave's velocity.
The gauge-Miura correspondence induces \emph{two solutions
of the KdV hierarchy}, which can be shown to have the form
\be\label{u_sol}
\u^{\pm} = \pa_x \eta^{\pm},\qquad
\eta^{\pm} = \v_0^2 x + c_{\v_0} t_{N} -2 \pa_x \log \tau^{\pm},
\ee
where the constant $c_{\v_0}$ only depends on the vacuum $\u_0 = \v_0^2$ and 
is model dependent, obtained to satisfy, e.g.,
relation \eqref{mt2} in the case of model \eqref{kdv_neg1} --- in this case
$c_{\v_0} = 2/\v_0$. When $\v_0=0$ the term $c_{\v_0}$ is absent
from this formula. Note that 
$\u^{\pm} = \u_0 - 2 \pa_x^2 \log \tau^{\pm}$ has
the well-known form of  ``Hirota's ansatz'' for the KdV equation.

\subsection{Complex KdV}
There are  some useful transformations worth noticing.
Eqs.~\eqref{kdv0} and \eqref{mkdv0} are known as \emph{defocusing}
KdV and mKdV, respectively; 
one can freely change $t\mapsto -t$ as well. 
Given $\u$ obeying eq.~\eqref{kdv0}, the transformation 
$\u \mapsto  -\u$  obeys the \emph{focusing} KdV, 
$4\pa_{t} \u - \pa_x^3 \u - 6 \u \pa_x \u = 0$. There is no qualitative
change in behavior except  for a sign shift.
The  mKdV \eqref{mkdv0} and sinh-Gordon \eqref{sg_eq} equations are
invariant under  $\v \mapsto -\v$.
Thus, to obtain the \emph{focusing} mKdV, 
$4 \pa_t \v - \pa_x^3 \v - 6 \v^2 \pa_x \v = 0$,
one needs to change
$\v \mapsto i \v$,  in which case
the sinh-Gordon becomes the sine-Gordon model,
$\pa_t \pa_x \phi = 2 \sin(2\phi) $. There is now a qualitative
change in the solutions' behavior. Moreover, such a purely imaginary
mKdV solution generates a \emph{complex} KdV solution via gauge-Miura.
More generally, suppose we have a complex solution $\v = \v_{R} + i \v_{I}$
of the mKdV eq.~\eqref{mkdv0}. Such components
satisfy a coupled  mKdV system
%\begin{subequations}\label{mkdv_sys}
%\begin{align}
%4 \pa_t \v_R  - \pa_x^3 \v_R + 6 \v_R^2 \pa_x \v_R 
%- 6 \pa_x\big( \v_R \v_I^2  \big) &= 0, \\
%4 \pa_t \v_I  - \pa_x^3 \v_{I} - 6 \v_I^2 \pa_x \v_I 
%+ 6 \pa_x\big( \v_R^2 \v_I \big) &= 0 ,
%\end{align}
%\end{subequations}
and play a completely \emph{symmetric} role.
However, a  complex solution $\u = \u_R + i \u_I$ of the KdV eq.~\eqref{kdv0} obeys
\begin{subequations}\label{kdv_sys}
\begin{align}
4 \pa_t \u_R - \pa_x^3 \u_R + 6 \u_R \pa_x \u_R &= 6 \u_I \pa_x \u_I, 
\label{kdv_sys1} \\
4 \pa_t \u_I - \pa_x^3 \u_I + 6 \pa_x( \u_R \u_I) &= 0,
\end{align}
\end{subequations}
whose components are  \emph{not  symmetric}. $\u_I$ acts as a source 
in the 1st equation, which  is a perturbed KdV equation.
This system corresponds to a particular case of the coupled
KdV system proposed in \cite{Brazhnyi_2005} in relation to 
Bose-Einstein condensates, obtained as a limit of 
coupled nonlinear Schr\" odinger
equations. One expects that such a system may capture 
an approximate dynamics of the latter \cite{Ankiewicz_2019}. 
As we illustrate below, this system is able to describe rich
nonlinear phenomena, including extreme events such as ``rogue waves.''

\subsection{The zoo of solitons and rogue waves} 
\begin{figure}[t!]
\centering
\begin{tabular}{p{0.0cm} p{2.2cm}  p{2.4cm} p{2.2cm} p{2.2cm} p{2.2cm} p{2.2cm} }
& $\phi$ (\mkdv) & $\v$ (\mkdv) & $\eta^+$ (\kdv) & $\u^{+}$ (\kdv) & $\eta^-$ (\kdv) & $\u^-$ (\kdv) 
\end{tabular}\\
\includegraphics[width=\textwidth]{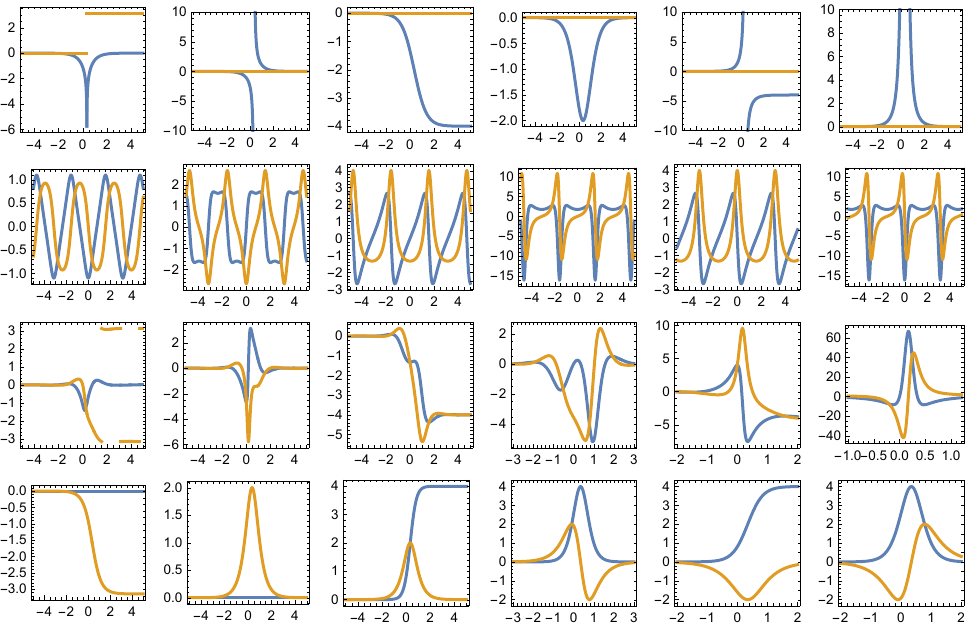}
\vspace{-2em}
\caption{\label{fig_sol1}
Zero vacuum, $\v_0 = \u_0 = 0$, ``1-soliton'' type of solutions.
We plot the fields in eqs.~\eqref{v_sol}--\eqref{u_sol} (columns) against $x$
for $t=0$. Blue line is the real part, orange line is the imaginary part.
\underline{\emph{1st row:}} Real (singular) mKdV solution with $\kappa = 1$
generates a KdV kink ($\eta^+$), a dark soliton ($\u^+$), 
and a peakon ($\u^-$);
changing $\u^\pm \to -\u^\pm$ yields a ``focusing'' KdV solution. 
\underline{\emph{2nd row:}} Periodic mKdV solutions   ($\kappa = i$).
Both $\u^{\pm}$ are called positons. 
\underline{\emph{3rd row:}}  Localized waves 
with complex  $\kappa = 1 + i $. $\u^-$ is a negaton (we shifted
$\u \mapsto -\u$) 
 and is reminiscent of the Peregrine soliton.
\underline{\emph{4th row:}} Purely imaginary mKdV solution ($\kappa = 1$, $f \mapsto f + i \pi /2$),
i.e., a kink and a typical soliton, generates
coupled kinks and solitons of focusing KdV.
}
\end{figure}

Through the formulas \eqref{v_sol}--\eqref{u_sol} we can generate
many KdV solutions --- including complex solutions --- from the 
mKdV tau functions.
In fig.~\ref{fig_sol1} we set  $n=1$ and classify several 
of such elementary solutions.
In the 1st row a \emph{real but singular} mKdV solution 
generates a \emph{KdV kink} for $\eta^+$,
a \emph{soliton or dark-soliton} for $\u^+$, as well as a \emph{peakon} 
for $\u^-$.
The 2nd row shows periodic solutions in space and time,
obtained with a purely imaginary wave number. The KdV solutions
in this case are known as \emph{positons} \cite{Hu_2006}.
The 3rd  row shows localized waves known as KdV \emph{negatons} \cite{Hu_2006}.
In particular,  $\u^-$ has  abnormal large amplitude and can be seen as a 
rogue wave of system \eqref{kdv_sys} (with $\u \mapsto -\u$); this solution may become singular at specific instants of time.
Interestingly, the real component is reminiscent of the 
Peregrine soliton \cite{Kibler_2010}, although over a zero 
background and this wave is persistent.
In the 4th row we add a phase $\theta = i \pi  / 2$ to the
dispersion relation.
This generates purely imaginary mKdV solutions, i.e.,
real solutions of the sine-Gordon or focusing
mKdV. Their KdV counterparts consist of \emph{complex} coupled solutions
involving \emph{kinks and solitons} (or dark-solitons by sign shift). Even in this case, KdV solitons tend to have larger amplitudes 
than mKdV ones, even if they were generated from the latter.

\begin{figure}[t!]
\centering
\begin{tabular}{p{0.0cm} p{2.2cm}  p{2.4cm} p{2.2cm} p{2.2cm} p{2.2cm} p{2.2cm} }
& $\phi$ (\mkdv) & $\v$ (\mkdv) & $\eta^+$ (\kdv) & $\u^{+}$ (\kdv) & $\eta^-$ (\kdv) & $\u^-$ (\kdv) 
\end{tabular}\\
\includegraphics[width=\textwidth]{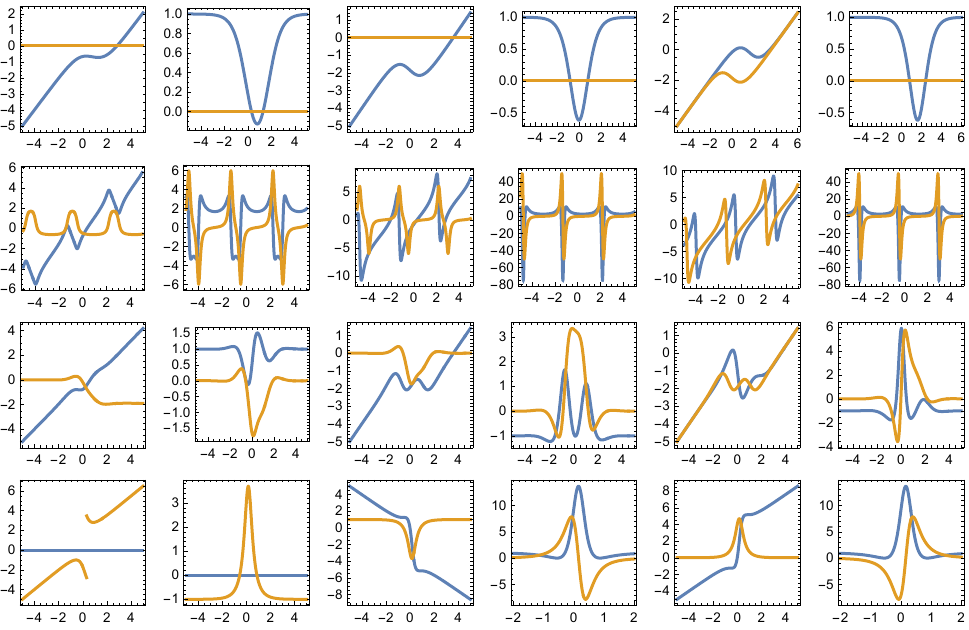}
\vspace{-2em}
\caption{\label{fig_sol2}
Nonzero vacuum, $\v_0 \ne 0$ and $\u_0 = \v_0^2$, ``1-soliton'' type of solutions.
We plot the fields in eqs.~\eqref{v_sol}--\eqref{u_sol} (columns) against $x$
for $t=0$. Blue line is the real part, orange line is the imaginary part.
\underline{\emph{1st row:}} $\kappa = 0.9$, $\v_0=1$.
Real mKdV dark soliton generates real KdV dark soliton.
\underline{\emph{2nd row:}} Periodic mKdV solutions,
$\kappa = 0.9 i$, $\v_0=1$.
Both $\u^{\pm}$ are positons over a nonzero  background. 
\underline{\emph{3rd row:}}  Localized waves, $\kappa = 0.9(1 + i) $, $\v_0=1$. For $\u^\pm$ we shifted the sign, yielding a large 
focusing KdV wave. 
\underline{\emph{4th row:}} Purely imaginary mKdV solution, $\kappa = 0.9$, $f \mapsto f + i \pi /2$, $\v_0 = i$.
This generates coupled KdV dark solitons and kinks. $\v$ has a large amplitude, and $\u^{\pm}$ even more so (they have the profile
of a Peregrine soliton).
} 
\end{figure}

In Fig.~\ref{fig_sol2} we generate a similar classification
but with a \emph{constant background} $\v_0 \ne 0$.
In the 1st row we have real mKdV dark solitons generating real KdV dark
solitons. The 2nd row have similar positons as in the previous figure,
but over a constant background.
The 3rd row shows KdV negatons over a constant background; the
amplitude of $\u^-$ is  significantly larger than   
their mKdV analogues.
The 4th row illustrates how focusing mKdV solutions 
generates large amplitude solitons for  complex KdV systems, including
system \eqref{kdv_sys}; this again has similar shape as the
Peregrine soliton \cite{Kibler_2010}. Even the mKdV dark soliton has 
relatively large amplitude.

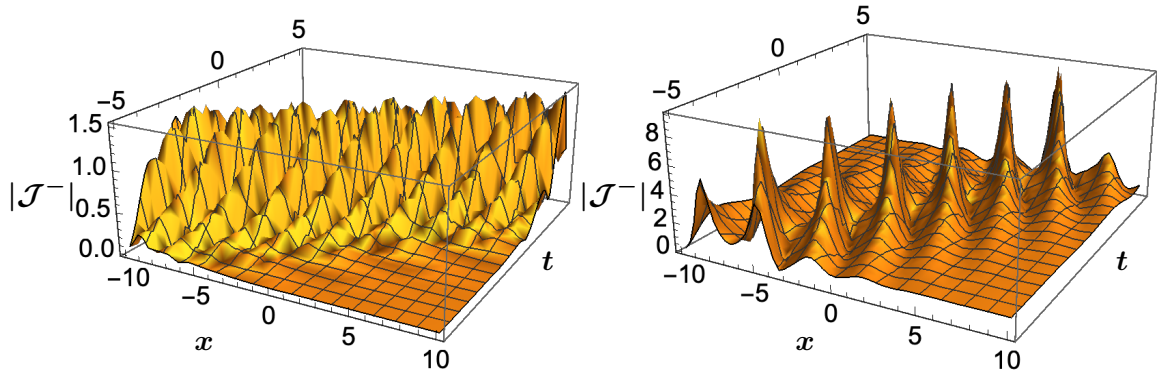
\begin{figure}[t]\centering
\includegraphics[width=.42\textwidth]{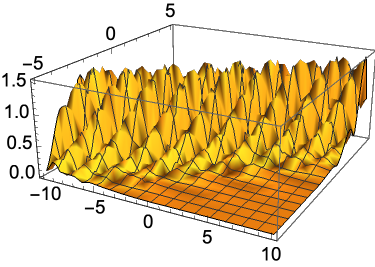}\qquad
\includegraphics[width=.42\textwidth]{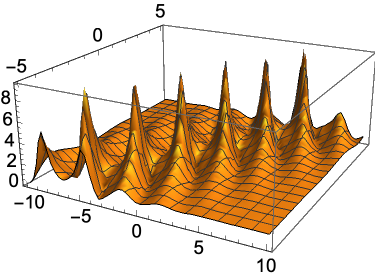}
\put(-360,10){$\bm{x}$}
\put(-135,10){$\bm{x}$}
\put(-230,40){$\bm{t}$}
\put(-15,40){$\bm{t}$}
\put(-430,65){$\bm{|\u^-|}$}
\put(-216,65){$\bm{|\u^-|}$}
\vspace{-.5em}
\caption{\label{fig_breather}
Complex KdV breathers.
``2-soliton'' solution (see eqs.~\eqref{v_sol}--\eqref{u_sol})
with $\kappa_{1,2} = 0.2 \pm 0.95 i$  
and a phase $i \pi /2$ in $f$.
These come from purely imaginary solutions of the mKdV hierarchy.
\underline{\emph{Left:}}
Zero vacuum, $\v_0 = 0$. 
\underline{\emph{Right:}}
Constant vacuum, $\v_0 = i$;
note the large amplitude of this breather, which is a rogue wave.
}
\end{figure}

In Fig.~\ref{fig_breather} we illustrate \emph{complex KdV breathers}
of system \eqref{kdv_sys}.
These are obtained from ``2-soliton'' ($n=2$) mKdV tau functions with complex
conjugate wave numbers.
Over a zero background, such breathers have a typical amplitude, however
over a nonzero background they have large amplitudes and provide a model for
rogue waves, as illustrated in the right plot.

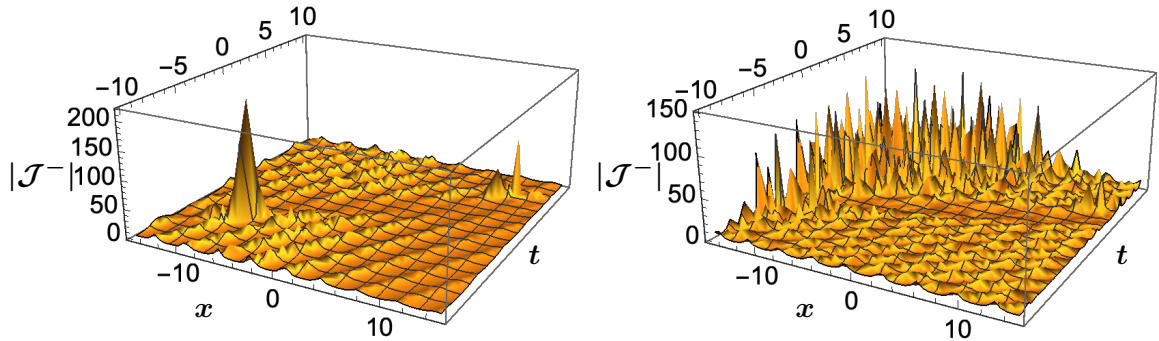
\begin{figure}[t]\centering
\includegraphics[width=.42\textwidth]{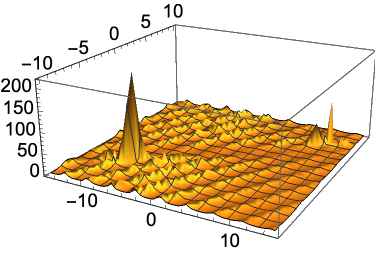}\qquad
\includegraphics[width=.42\textwidth]{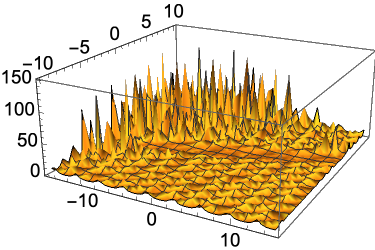}
\put(-360,10){$\bm{x}$}
\put(-135,10){$\bm{x}$}
\put(-235,30){$\bm{t}$}
\put(-15,30){$\bm{t}$}
\put(-430,60){$\bm{|\u^-|}$}
\put(-210,60){$\bm{|\u^-|}$}
\vspace{-.5em}
\caption{\label{fig_rogue}
Complex KdV rogue waves.
``4-soliton'' solution (see eqs.~\eqref{v_sol}--\eqref{u_sol});
$\kappa_1 = 0.1 + 0.1 i$, $\kappa_{2,3} = 0.3\pm 0.9 i$,
$\kappa_4 = 0.91 i$,  and a phase $i \pi /2$ in $f$.
\underline{\emph{Left:}}
Zero vacuum, $\v_0 = 0$. Note the large localized
rogue wave. 
\underline{\emph{Right:}}
Constant vacuum, $\v_0 = 1$.
Note the extremely violent and high amplitude waves.
By changing parameters it is possible to create localized waves
as in the left figure but with much larger amplitude.
}
\end{figure}

In Fig.~\ref{fig_rogue} we consider a ``4-soliton'' solution ($n=4$)
of the mKdV hierarchy, which induces solutions to system
\eqref{kdv_sys}; the parameters are indicated in the caption.
We generate an extreme localized wave (left plot) even over a zero
vacuum. For the same parameters, however over a constant
vacuum $\v_0=1$, such waves model extremely violent phenomena (right plot)
with rogue waves scattered in several places.
By tweaking parameters, it is possible to obtain a localized rogue wave,
although with much larger amplitude  compared to the zero background case.  

\section{Conclusion}
We have constructed solutions of the (focusing/defocusing) 
KdV hierarchy in  terms of
tau functions of the mKdV hierarchy. Such solutions hold
for any model of the hierarchy by adapting the dispersion relation, including the specific complex KdV system \eqref{kdv_sys}.
We  have illustrated how rogue waves can be obtained from simple analytical
formulas.  Our examples are by no means exhaustive, i.e., 
by suitable combination
of elementary mKdV tau function one can construct
countless examples of extreme waves and model rich nonlinear phenomena.

\ack
JFG and AHZ thank CNPq
and FAPESP for support. YFA thanks FAPESP for financial support under grant 2021/00623-4 and 2022/13584-0. GVL thanks
CAPES (finance Code 001).

\vspace{-1em}

\section*{References}
\bibliography{biblio.bib}

\end{document}